\documentclass[12pt]{article}
\usepackage{color,amssymb,amsthm,amsmath,fullpage,graphicx,hyperref,float,bm,array,booktabs}

\usepackage{fancyhdr}
\usepackage{xcolor}
\definecolor{dark-blue}{rgb}{0,0,0.45}
\hypersetup{colorlinks, linkcolor={dark-blue}, citecolor={blue}, urlcolor={blue}}

\numberwithin{equation}{section}

\def\Z{\mathbb{Z}}
\def\C{\mathbb{C}}
\def\H{\mathbb{H}}
\def\Qe{Q_8}
\def\E{\mathcal{E}}
\def\il{\mathbin{\between}}

\DeclareMathOperator{\ord}{ord}

\definecolor{gold}{rgb}{0.85,0.66,0.0}

\newtheorem{theorem}{Theorem}[section]
\newtheorem{lemma}[theorem]{Lemma}

\newtheorem{proposition}[theorem]{Proposition}
\newtheorem{corollary}[theorem]{Corollary}
\theoremstyle{definition}
\newtheorem{example}[theorem]{Example}
\newtheorem{definition}[theorem]{Definition}
\newtheorem{construction}[theorem]{Construction}

\title{A Recursive Construction of Perfect Arrays of\\ Every Shape $2^{a_1}\times\cdots\times2^{a_d}$ over the Unit Quaternions}
\author{Sam Blake}
\date{15 November 2026}

\begin{document}

\maketitle

\begin{abstract}
We extend the recursive construction of perfect and odd perfect sequences over the simple unit
quaternions $\Qe=\{\pm1,\pm i,\pm j,\pm k\}$ of \cite{Q8tower} to arrays of any dimension. The
three operations of that construction act on one direction of an array at a time, and are valid
whenever the array is periodic in the remaining directions. Starting from a perfect array $P$ of
shape $M_2\times\cdots\times M_d$ they give perfect arrays of shape
$2^a\times M_2\times\cdots\times M_d$ for every $a\ge0$, with explicit index functions and with the
array orthogonality property for the divisor $2$ in the lifted direction; from $a=5$ on these
arrays are, in general, Kronecker products of a sequence with $P$ in neither order. Applied
direction by direction, the construction gives perfect arrays over $\Qe$ in every shape
$2^{a_1}\times\cdots\times2^{a_d}$, and in every shape $2^{a_1}\times\cdots\times2^{a_{d-1}}\times N$
for which a perfect sequence of length $N$ over $\Qe$ exists. The contrast with the binary alphabet
is sharp: a perfect binary array of shape $2^{a_1}\times\cdots\times2^{a_d}$ exists only when
$\sum a_k$ is even and $\max a_k\le\frac12\sum a_k+1$, by a theorem of Jedwab and Kraemer on
Menon difference sets in abelian $2$-groups, whereas over $\Qe$ there is no constraint on the shape. We
also show that, unlike the one-dimensional case, single arrays carrying a non-central twist do exist
in two dimensions, and determine where the one-dimensional non-existence argument breaks.
\end{abstract}

\section{Introduction}

A $d$-dimensional array over the simple unit quaternions $\Qe=\{\pm1,\pm i,\pm j,\pm k\}$ is
\emph{perfect} if all of its off-peak periodic autocorrelations vanish. Over the binary alphabet
$\{\pm1\}$, a nontrivial perfect array is equivalent to a Menon difference set in the abelian group
of its shape \cite[Theorem~3.1]{Jedwab92}, and for shapes $2^{a_1}\times\cdots\times2^{a_d}$ the
existence question is settled: a perfect binary array of that shape exists if and only if
$\sum a_k=2y\ge2$ is even and $a_k\le y+1$ for all $k$ \cite[Theorem~9.2(i)]{Jedwab92}, the
necessity being Kraemer's theorem that an abelian group of order $2^{2y}$ contains a Menon
difference set only if its exponent is at most $2^{y+1}$ \cite{Kraemer93}. The shapes $2^n\times2^n$ are thus all
available, but $128\times2$, $64\times4$, $32\times2\times2$ and in general any shape with one
side much longer than the rest are not, and no binary array has a side of odd length. Over roots
of unity the situation is similar in spirit, and Mow's conjecture \cite{Mow1993,Mow1996} bounds the
length of a perfect sequence over $n$ roots of unity by $n^2$. Over $\Qe$, where the alphabet is
fixed but non-commutative, perfect sequences of every length $2^t$ were constructed by Bright,
Kotsireas and Ganesh \cite{BKG20} and, with explicit index functions, in \cite{Q8tower}; perfect
arrays of unbounded size over $\Qe$ were obtained from Lee sequences by Barrera Acevedo and Jolly
\cite{BAJolly14}, and arrays of shape $2^n\times2^n$ by an index function in
\cite[\S8.4]{Blake2017}, verified there for $n<7$.

This paper shows that the recursive construction of \cite{Q8tower} lifts to arrays of any
dimension and removes the constraint on the shape entirely. The construction of \cite{Q8tower}
rests on three operations: an orthogonality lemma for twisted periodic extensions with distinct
commuting twists, two lifts which interleave twisted extensions to multiply the length by $4$, and
a doubling of \emph{$i$-pairs}, the quaternion analogue of Golay pairs, in which the sign of
Golay's construction becomes a right factor $j$. All three act on a sequence by interleaving and
twisting, and the observation of this paper is that each of them acts on \emph{one direction} of an
array, and is valid for arrays periodic in the remaining directions, by the same proofs. Starting
from any perfect array $P$ of shape $M'=M_2\times\cdots\times M_d$ over $\Qe$, the seed $(P,P)$
is an $i$-pair in the new direction, and the tower of \cite{Q8tower} runs in that direction
unchanged, giving perfect arrays of every shape $2^a\times M'$ with the array orthogonality
property (AOP) for the divisor $2$ in the lifted direction (Theorem~\ref{towerthm}). Applied
direction by direction, this gives perfect arrays over $\Qe$ of every shape
$2^{a_1}\times\cdots\times2^{a_d}$, and of every shape
$2^{a_1}\times\cdots\times2^{a_{d-1}}\times N$ for which a perfect sequence of length $N$ over $\Qe$
exists (Corollary~\ref{shapecor}).

We should say at once what is and is not new in this. The Kronecker product of perfect arrays over
$\Qe$ is perfect, with the sequence placed on either side (Proposition~\ref{kronprop}), so the
existence of all shapes $2^{a_1}\times\cdots\times2^{a_d}$ is already a consequence of the
sequences of \cite{Q8tower}, and the arrays of Theorem~\ref{towerthm} with at most $2^4$ along the
lifted direction are such products. What the tower adds is a family of perfect arrays which, from
$2^5$ on, are products in neither order (Section~\ref{towersec}), with the nested index functions of
\cite{Q8tower,BlakeTower} and the AOP in the lifted direction; and a framework, the twisted types
of \cite{BlakeTower} over a non-commutative alphabet, in which the dimension enters only through the
periodicity of the base. In Section~\ref{notypesec} we show that this framework behaves differently
in dimension two from dimension one: single arrays of non-central type, which do not exist for
sequences of length $M\not\equiv0\pmod4$ \cite[Proposition~5.1]{Q8tower}, exist already in shape
$2\times2$, and the reduction to a quaternary array survives in every dimension while the parity
argument does not.

All of the results below were found by computer before being proved; the program
\texttt{nd.py} accompanying the paper constructs every array mentioned and checks its
autocorrelations exactly.

\section{Definitions}

\subsection{Quaternions and arrays}
We use the notation of \cite{Q8tower}: $\H$ is the real quaternions with $i^2=j^2=k^2=ijk=-1$,
$q^*$ is the conjugate, $(pq)^*=q^*p^*$, $\Qe=\{\pm1,\pm i,\pm j,\pm k\}$ with centre $\{\pm1\}$,
and two elements of $\Qe$ commute if and only if one is central or both lie in the same cyclic
subgroup. We write $q=\alpha+\beta j$ with $\alpha,\beta\in\C$, and call the entries of an array
lying in $\{\pm1,\pm i\}$ \emph{complex} and those in $\{\pm j,\pm k\}$ \emph{imaginary}, with
$N_C$ and $N_J$ for their numbers.

A $d$-dimensional array $A=[A_{\mathbf n}]$ of shape $M_1\times\cdots\times M_d$ over $\Qe$ is
indexed by $\mathbf n=(n_1,\dots,n_d)$ with $n_k$ taken modulo $M_k$. We write $\mathbf n=(n_1,
\mathbf n')$ with $\mathbf n'=(n_2,\dots,n_d)$, and $M'=M_2\times\cdots\times M_d$ for the shape of
the remaining directions; a sequence is the case $d=1$, and an array of shape $1\times M'$ is the
same thing as an array of shape $M'$. The periodic cross-correlation of two arrays $A,B$ of a common
shape $L_1\times\cdots\times L_d$ for the shift $\boldsymbol\tau$ is
\[
\theta_{A,B}(\boldsymbol\tau)=\sum_{\mathbf n}A_{\mathbf n}\,B_{\mathbf n+\boldsymbol\tau}^*,
\]
the sum over one period, and $\theta_A=\theta_{A,A}$ is the periodic autocorrelation. Two arrays
are \emph{orthogonal} if $\theta_{A,B}$ vanishes identically. As the entries are units,
$\theta_A(\mathbf0)$ is the number of entries and $\theta_A(-\boldsymbol\tau)=\theta_A(\boldsymbol
\tau)^*$; for a unit $v$, $\theta_{Av}=\theta_A$ and $\theta_{vA}=v\,\theta_A\,v^*$. An array is
\emph{perfect} if $\theta_A(\boldsymbol\tau)=0$ for all $\boldsymbol\tau\ne\mathbf0$. Kuznetsov
\cite{Kuz09} proved that a quaternion sequence is perfect for the correlation above if and only if
it is perfect for the correlation with the conjugate on the first factor, and the same proof applies
to arrays, so nothing depends on this choice.

\subsection{Twisting, types and interleaving in one direction}
All operations below act on direction $1$ and leave the remaining directions periodic.

\begin{definition}\label{typedef}
Let $A$ have shape $M_1\times M'$ and $u\in\Qe$. The \emph{right $u$-twisted extension of $A$ in
direction $1$} is the array $\E_u[A]$ of shape $\ord(u)M_1\times M'$ with
\[
\E_u[A]_{(qM_1+m,\,\mathbf n')}=A_{(m,\mathbf n')}\,u^q,\qquad0\le m<M_1,
\]
so that $\E_u[A]_{\mathbf n+(M_1,\mathbf0)}=\E_u[A]_{\mathbf n}\,u$. The array $A$ has \emph{type
$u$ in direction $1$} if $\theta_{\E_u[A]}(\boldsymbol\tau)=0$ for every $\boldsymbol\tau$ outside
the set $\{(qM_1,\mathbf0)\}$.
\end{definition}

Type $1$ is perfection, and type $-1$ we call \emph{odd perfect in direction $1$}: the array
$[A;-A]$ obtained by continuing $A$ nega-periodically in direction $1$ is perfect off the shifts
$(qM_1,\mathbf0)$. The twisted autocorrelation of $A$ is
$\vartheta^u_A(\boldsymbol\tau)=\sum_{0\le m<M_1,\,\mathbf n'}A_{(m,\mathbf n')}\,
(\E_u[A]_{(m,\mathbf n')+\boldsymbol\tau})^*$, and exactly as in \cite[Lemma~2.1]{Q8tower},
computing over any period $L_1\times M'$ with $\ord(u)M_1\mid L_1$,
\begin{equation}\label{Teq}
\theta_{\E_u[A]}(\boldsymbol\tau)=\frac{L_1}{M_1}\,\vartheta^u_A(\boldsymbol\tau),
\end{equation}
because $\E_u[A]_{\mathbf n}(\E_u[A]_{\mathbf n+\boldsymbol\tau})^*=A_{(m,\mathbf n')}u^q(u^q)^*
(\E_u[A]_{(m,\mathbf n')+\boldsymbol\tau})^*$ does not depend on $q$. For $u^2=-1$ one has, over
the period $4M_1\times M'$, $\theta(\boldsymbol\tau+(2M_1,\mathbf0))=-\theta(\boldsymbol\tau)$ and
$\theta((2M_1,\mathbf0)-\boldsymbol\tau)=-\theta(\boldsymbol\tau)^*$, as in
\cite[Lemma~2.2]{Q8tower}. Left multiplication by a unit preserves every type, right
multiplication by $v$ sends type $u$ to type $vuv^*$ since $\E_u[Av]=\E_{vuv^*}[A]\,v$, and an
automorphism of $\Qe$ applied entrywise permutes the non-central types transitively; we therefore
take the non-central twist to be $i$ throughout.

Twists in several directions at once must commute: if $\E$ satisfied
$\E_{\mathbf n+(M_1,\mathbf0)}=\E_{\mathbf n}u$ and $\E_{\mathbf n+(0,M_2,\mathbf0)}=\E_{\mathbf n}
v$ then $\E_{\mathbf n+(M_1,M_2,\mathbf 0)}$ would equal both $\E_{\mathbf n}uv$ and
$\E_{\mathbf n}vu$. Since the abelian subgroups of $\Qe$ are cyclic of order at most $4$, a vector
of twists $(u_1,\dots,u_d)$ lies in $\{1,-1,u,-u\}^d$ for a single $u$; the types of a
$d$-dimensional array over $\Qe$ are, in effect, the types of \cite{BlakeTower} over $\Z_4$
attached to one non-central element. In this paper only direction $1$ carries a twist.

\begin{definition}
Let $B,C$ have a common shape $N\times M'$. Their \emph{interleaving in direction $1$} is the array
$B\il C$ of shape $2N\times M'$ with $(B\il C)_{(2n+r,\mathbf n')}=B_{(n,\mathbf n')}$ for $r=0$
and $C_{(n,\mathbf n')}$ for $r=1$.
\end{definition}

The reindexing that proves \cite[Theorem~1.1.9]{Blake2017} gives, with the correlations on the
right computed over $N\times M'$,
\begin{align}
\theta_{B\il C}(2\tau_1,\boldsymbol\tau')&=\theta_B(\tau_1,\boldsymbol\tau')+\theta_C(\tau_1,
\boldsymbol\tau'),\label{ileven}\\
\theta_{B\il C}(2\tau_1+1,\boldsymbol\tau')&=\theta_{B,C}(\tau_1,\boldsymbol\tau')+
\theta_{C,B}(\tau_1+1,\boldsymbol\tau'),\label{ilodd}
\end{align}
and interleaving commutes with twisting in direction $1$: $\E_u[B\il C]=\E_u[B]\il\E_u[C]$. The
array $B\il C$ has the \emph{array orthogonality property for the divisor $2$ in direction $1$}
if $B$ and $C$ are orthogonal and $\theta_B+\theta_C$ vanishes off $\mathbf0$; by
\eqref{ileven}--\eqref{ilodd} it is then perfect. This is the AOP of \cite[\S1.1]{Blake2017} for
the decomposition of direction $1$ by the divisor $2$.

\subsection{Products}
For a sequence $x$ of length $N$ and an array $P$ of shape $M'$ let $x\otimes P$ and
$P\otimes x$ be the arrays of shape $N\times M'$ with
\[
(x\otimes P)_{(a,\mathbf n')}=x_a\,P_{\mathbf n'},\qquad(P\otimes x)_{(a,\mathbf n')}=
P_{\mathbf n'}\,x_a .
\]
Over a commutative alphabet these coincide; over $\Qe$ they differ, and both behave well.

\begin{proposition}\label{kronprop}
Let $P$ be a perfect array of shape $M'$ over $\Qe$. If $x$ is a perfect sequence, then
$x\otimes P$ and $P\otimes x$ are perfect; if $x$ is odd perfect, then $x\otimes P$ and
$P\otimes x$ are odd perfect in direction $1$.
\end{proposition}
\begin{proof}
For $x\otimes P$ and the shift $(s,\boldsymbol\tau')$,
\[
\theta_{x\otimes P}(s,\boldsymbol\tau')=\sum_a\sum_{\mathbf n'}x_aP_{\mathbf n'}
P^*_{\mathbf n'+\boldsymbol\tau'}x^*_{a+s}=\sum_ax_a\,\theta_P(\boldsymbol\tau')\,x^*_{a+s}.
\]
For $\boldsymbol\tau'\ne\mathbf0$ the inner factor is $0$; for $\boldsymbol\tau'=\mathbf0$ it is
the real scalar $|M'|$, which commutes with everything, so the sum is $|M'|\theta_x(s)$, zero for
$s\ne0$. For $P\otimes x$ the inner sum is $\sum_ax_ax^*_{a+s}$, again a real scalar or zero. The
odd perfect case is the same computation for $\E_{-1}[x]\otimes P=\E_{-1}[x\otimes P]$.
\end{proof}

Thus any perfect sequence over $\Qe$ of length $N$ and any perfect array of shape $M'$ give a
perfect array of shape $N\times M'$, and the shapes of Corollary~\ref{shapecor} below follow
already from the sequences of \cite{Q8tower}. The tower of Section~\ref{towersec} produces arrays
that are not of this form.

\section{Orthogonality and the two lifts in one direction}\label{liftsec}

\begin{lemma}[orthogonality]\label{orthlem}
Let $A,B$ have shape $M_1\times M'$, let $u,v\in\Qe$ commute with $u\ne v$, and let $L_1$ be a
common multiple of $\ord(u)M_1$ and $\ord(v)M_1$. Then, over the period $L_1\times M'$,
$\theta_{\E_u[A],\E_v[B]}(\boldsymbol\tau)=0$ for every $\boldsymbol\tau$.
\end{lemma}
\begin{proof}
Writing $\mathbf n=(qM_1+m,\mathbf n')$, the summand is $A_{(m,\mathbf n')}\,u^q(v^q)^*\,
(\E_v[B]_{(m,\mathbf n')+\boldsymbol\tau})^*=A_{(m,\mathbf n')}(uv^*)^q(\E_v[B]_{(m,\mathbf n')
+\boldsymbol\tau})^*$, and $\sum_{q=0}^{L_1/M_1-1}(uv^*)^q=0$ since $1\ne uv^*\in\Qe$ has order
dividing $L_1/M_1$ \cite[Lemma~8.0.2]{Blake2017}.
\end{proof}

\begin{definition}\label{pairdef}
Arrays $A_0,A_1$ of shape $M_1\times M'$ form an \emph{$i$-pair in direction $1$} if, over the
period $4M_1\times M'$,
\[
\theta_{\E_i[A_0]}(\boldsymbol\tau)+\theta_{\E_{-i}[A_1]}(\boldsymbol\tau)=0\qquad
\text{for all }\boldsymbol\tau\notin\{(2qM_1,\mathbf0)\}.
\]
\end{definition}

By the symmetries following \eqref{Teq} the condition reduces to
$\vartheta^i_{A_0}+\vartheta^{-i}_{A_1}=0$ for $0<\tau_1\le M_1$ and all $\boldsymbol\tau'$, and
at $\boldsymbol\tau=(M_1,\mathbf0)$ it reads $N_C(A_0)-N_J(A_0)=N_C(A_1)-N_J(A_1)$, as in
\cite[(4.2)]{Q8tower}. The following seed is what makes the construction run in a new direction.

\begin{lemma}[seed]\label{seedlem}
Let $P$ be a perfect array of shape $M'$, regarded as an array of shape $1\times M'$. Then $P$
is perfect and odd perfect in direction $1$, $[P;iP]=[1,i]\otimes P$ is perfect, $[P;P]=[1,1]
\otimes P$ is odd perfect in direction $1$, and $(P,P)$ is an $i$-pair in direction $1$.
\end{lemma}
\begin{proof}
The first two statements say that $\E_{\pm1}[P]$ is perfect off $\{(q,\mathbf0)\}$, which is the
perfection of $P$ in the remaining directions; the next two are Proposition~\ref{kronprop} with the
perfect sequence $[1,i]$ and the odd perfect sequence $[1,1]$. For the pair, with $M_1=1$,
$\vartheta^{i}_P(\tau_1,\boldsymbol\tau')=\sum_{\mathbf n'}P_{\mathbf n'}(i^{\tau_1})^*
P^*_{\mathbf n'+\boldsymbol\tau'}$ and $\vartheta^{-i}_P(\tau_1,\boldsymbol\tau')=\sum_{\mathbf n'}
P_{\mathbf n'}((-i)^{\tau_1})^*P^*_{\mathbf n'+\boldsymbol\tau'}$. For $\tau_1=1$ the two unit
factors are $i^*$ and $i$, which cancel; for $\tau_1=2$ both are $-1$, and the sum is
$-2\theta_P(\boldsymbol\tau')$, which vanishes for $\boldsymbol\tau'\ne\mathbf0$, while
$(2,\mathbf0)$ is excluded.
\end{proof}

\begin{theorem}[the two lifts]\label{liftthm}
Let $A_0,A_1$ have shape $M_1\times M'$.
\begin{enumerate}
\item[(i)] If $(A_0,A_1)$ is an $i$-pair in direction $1$, then $\E_i[A_0]\il\E_{-i}[A_1]$, as an
array of shape $4M_1\times M'$, is odd perfect in direction $1$.
\item[(ii)] If $A_0$ is perfect and $A_1$ is odd perfect in direction $1$, then
$\E_1[A_0]\il\E_{-1}[A_1]$, as an array of shape $4M_1\times M'$, is perfect and has the AOP for
the divisor $2$ in direction $1$.
\end{enumerate}
\end{theorem}
\begin{proof}
Identical to \cite[Theorem~4.3]{Q8tower}: with $u=i$ or $u=1$ and $w=u^2$, the components
$c_0=\E_u[A_0]$, $c_1=\E_{-u}[A_1]$ satisfy $c_{\mathbf n+(2M_1,\mathbf0)}=c_{\mathbf n}w$, so the
interleaving is $\E_w[Y]$ with $Y$ its first $4M_1$ rows; the even shifts \eqref{ileven} vanish by
the pair condition (in case (ii), by \cite[Lemma~4.2]{Q8tower}, which only uses the symmetries
following \eqref{Teq}), and the odd shifts \eqref{ilodd} vanish by Lemma~\ref{orthlem} since $u$
and $-u$ commute and differ. In case (ii) the same two facts are the AOP.
\end{proof}

\section{Doubling in one direction}\label{doublesec}

\begin{theorem}[doubling]\label{doublethm}
Let $(A_0,A_1)$ be an $i$-pair in direction $1$ of shape $M_1\times M'$ and let
$v\in\{\pm j,\pm k\}$. Then $(A_0\il A_1v,\ A_0v\il A_1)$ is an $i$-pair in direction $1$ of
shape $2M_1\times M'$.
\end{theorem}
\begin{proof}
As $v$ anticommutes with $i$, $\E_i[A_1v]=\E_{-i}[A_1]v$ and $\E_{-i}[A_0v]=\E_i[A_0]v$, so
with $c_0=\E_i[A_0]$ and $c_1=\E_{-i}[A_1]$ the twisted extensions of the two new arrays are
$c_0\il c_1v$ and $c_0v\il c_1$. The even shifts of each are
$\theta_{c_0}+\theta_{c_1}$ by \eqref{ileven} and $\theta_{cv}=\theta_c$, and vanish off
$\{(2qM_1,\mathbf0)\}$ by hypothesis. The odd shifts are, by \eqref{ilodd},
\[
\sum c_{0,\mathbf n}\,v^*\,c^*_{1,\mathbf n+\boldsymbol\sigma}+\sum c_{1,\mathbf n}\,v\,
c^*_{0,\mathbf n+\boldsymbol\sigma+(1,\mathbf 0)}
\quad\text{and}\quad
\sum c_{0,\mathbf n}\,v\,c^*_{1,\mathbf n+\boldsymbol\sigma}+\sum c_{1,\mathbf n}\,v^*\,
c^*_{0,\mathbf n+\boldsymbol\sigma+(1,\mathbf0)},
\]
which cancel term by term because $v+v^*=0$.
\end{proof}

The proof is that of \cite[Theorem~6.1]{Q8tower} with a second index carried along; nothing in it
sees the remaining directions, which is the whole point.

\section{The tower in one direction}\label{towersec}

Fix a perfect array $P$ of shape $M'$ over $\Qe$. For $t\ge0$, $r\in\{0,1\}$ and $0\le a<2^t$
with binary digits $a=\sum_{\ell<t}b_\ell2^\ell$, let $w_t(r,a)$ be the number of changes in the
bit string $(r,b_0,\dots,b_{t-1})$, as in \cite[(7.1)]{Q8tower}, and let $y^{(t)}$ be the odd
perfect sequence of length $2^{t+2}$ of \cite[Construction~7.2]{Q8tower},
\[
y^{(t)}_{2a+r}=j^{\,w_t(r,\,a\bmod2^t)}\,i^{\,(1-2r)\lfloor a/2^t\rfloor}.
\]

\begin{construction}\label{pairconstr}
For $t\ge0$ and $r\in\{0,1\}$ let $A^{(t)}_r$ be the array of shape $2^t\times M'$ with
$A^{(t)}_{r,(a,\mathbf n')}=P_{\mathbf n'}\,j^{\,w_t(r,a)}$.
\end{construction}

\begin{construction}\label{oddconstr}
For $t\ge0$ let $Y^{(t)}=P\otimes y^{(t)}$, the array of shape $2^{t+2}\times M'$ with
$Y^{(t)}_{(n,\mathbf n')}=P_{\mathbf n'}\,y^{(t)}_n$.
\end{construction}

\begin{construction}\label{perfconstr}
Let $p^{(0)}=P$, $p^{(1)}=[P;iP]$, $q^{(0)}=P$, $q^{(1)}=[P;P]$ and $q^{(t+2)}=Y^{(t)}$. For
$a\ge2$ let $p^{(a)}$ be the array of shape $2^a\times M'$ with
\[
p^{(a)}_{(2b+r,\mathbf n')}=\begin{cases}p^{(a-2)}_{(b\bmod2^{a-2},\,\mathbf n')},&r=0,\\[2pt]
q^{(a-2)}_{(b\bmod2^{a-2},\,\mathbf n')}\,(-1)^{\lfloor b/2^{a-2}\rfloor},&r=1,\end{cases}
\qquad0\le b<2^{a-1}.
\]
\end{construction}

\begin{theorem}\label{towerthm}
Let $P$ be a perfect array of shape $M'$ over $\Qe$. For every $t\ge0$,
$(A^{(t)}_0,A^{(t)}_1)$ is an $i$-pair in direction $1$ and $Y^{(t)}$ is odd perfect in direction
$1$. For every $a\ge0$, $p^{(a)}$ is a perfect array of shape $2^a\times M'$ over $\Qe$, with the
AOP for the divisor $2$ in direction $1$ when $a\ge2$.
\end{theorem}
\begin{proof}
Since $w_t(s,2a+r)=[s\ne r]+w_{t-1}(r,a)$, we have $A^{(t)}_{s,(2a+r,\mathbf n')}=
A^{(t-1)}_{r,(a,\mathbf n')}j^{[s\ne r]}$, i.e.\ $A^{(t)}_0=A^{(t-1)}_0\il A^{(t-1)}_1j$ and
$A^{(t)}_1=A^{(t-1)}_0j\il A^{(t-1)}_1$, which is Theorem~\ref{doublethm} with $v=j$; the seed
$A^{(0)}_0=A^{(0)}_1=P$ is an $i$-pair by Lemma~\ref{seedlem}. With $a=q2^t+a'$,
$\E_i[A^{(t)}_0]_{(a,\mathbf n')}=P_{\mathbf n'}j^{w_t(0,a')}i^q$ and
$\E_{-i}[A^{(t)}_1]_{(a,\mathbf n')}=P_{\mathbf n'}j^{w_t(1,a')}(-i)^q$, so
$Y^{(t)}=\E_i[A^{(t)}_0]\il\E_{-i}[A^{(t)}_1]$ and Theorem~\ref{liftthm}(i) applies. Finally
$p^{(0)},p^{(1)}$ are perfect and $q^{(0)},q^{(1)}$ odd perfect in direction $1$ by
Lemma~\ref{seedlem}, $p^{(a)}=\E_1[p^{(a-2)}]\il\E_{-1}[q^{(a-2)}]$ by definition, and
Theorem~\ref{liftthm}(ii) gives the induction step.
\end{proof}

Note that the $i$-pairs are $P\otimes x^{(t)}_r$ for the sequences $x^{(t)}_r=[j^{w_t(r,a)}]$ of
\cite[Construction~7.1]{Q8tower}, and the odd perfect arrays are $P\otimes y^{(t)}$: for these the
tower only reproduces Proposition~\ref{kronprop}, with the sequence on the right. The perfect arrays
are different. For $a\le4$ the recursion of Construction~\ref{perfconstr} only involves $q^{(0)}$,
$q^{(1)}$ and $q^{(2)}=P\otimes[1,1,1,-1]$, all of which are left products as well since their
sequence factors are real, and one checks that $p^{(a)}=p_{\mathrm{seq}}^{(a)}\otimes P$ for the
perfect sequences $p^{(a)}_{\mathrm{seq}}$ of \cite[Construction~7.3]{Q8tower}. From $a=5$ the
array $q^{(3)}=P\otimes y^{(1)}$ with $y^{(1)}=[1,j,j,1,i,k,-k,-i]$ enters, and $p^{(a)}$ mixes
rows of the form $P$, $iP$ with rows of the form $Pu$, $u\in\Qe$. For $P=[1,i]$ we have verified
that $p^{(a)}$, $5\le a\le7$, is not of the form $x\otimes P$ or $P\otimes x$ for any sequence
$x$, and the same holds for $P$ equal to the perfect sequences of lengths $8$ and $16$ of
\cite{Q8tower}; when $P$ is real, as for $P=[1,1,1,-1]$, every $p^{(a)}$ is a product.

\begin{example}\label{p45ex}
For $P=[1,i]$ the arrays $p^{(4)}$ and $p^{(5)}$, of shapes $16\times2$ and $32\times2$, have the
rows (read across, then down)
\[
p^{(4)}:\ \begin{array}{l}
(1,i),(1,i),(1,i),(1,i),(1,i),(i,-1),(-1,-i),(-i,1),\\
(1,i),(-1,-i),(1,i),(-1,-i),(1,i),(-i,1),(-1,-i),(i,-1);
\end{array}
\]
\[
p^{(5)}:\ \begin{array}{l}
(1,i),(1,i),(1,i),(j,k),(i,-1),(j,k),(1,i),(1,i),\\
(1,i),(i,-1),(-1,-i),(k,-j),(i,-1),(-k,j),(-1,-i),(-i,1),\\
(1,i),(-1,-i),(1,i),(-j,-k),(i,-1),(-j,-k),(1,i),(-1,-i),\\
(1,i),(-i,1),(-1,-i),(-k,j),(i,-1),(k,-j),(-1,-i),(i,-1).
\end{array}
\]
The first is $p^{(4)}_{\mathrm{seq}}\otimes[1,i]$; the second is not a product in either order.
Its rows $(j,k)=(1,i)j$ and $(k,-j)=(1,i)k$ come from $q^{(3)}=P\otimes y^{(1)}$, and its rows
$(i,-1)=i\,(1,i)$ from $p^{(1)}=[P;iP]$; no sequence multiplies $(1,i)$ into both from one side.
\end{example}

\begin{corollary}\label{shapecor}
Perfect arrays over $\Qe$ exist in every shape $2^{a_1}\times\cdots\times2^{a_d}$, $a_k\ge0$, and
in every shape $2^{a_1}\times\cdots\times2^{a_{d-1}}\times N$ for which a perfect sequence of length
$N$ over $\Qe$ exists; in particular for $N=6\cdot4^s$ \cite[Proposition~8.3]{Q8tower} and for the
lengths of the sequences of \cite{BAHall12}, which lie over $\{\pm1,\pm i,j\}\subset\Qe$. The arrays
obtained have the AOP for the divisor $2$ in the last direction lifted, whenever that direction has
length at least $4$.
\end{corollary}
\begin{proof}
Induction on $d$ with Theorem~\ref{towerthm}, starting from the sequence $[1]$ or from a perfect
sequence of length $N$ in the last direction: each application lifts one new direction to length
$2^{a}$ and leaves a perfect array of one more dimension as the base for the next. The AOP in the last
direction lifted is Theorem~\ref{towerthm}.
\end{proof}

We have verified Theorem~\ref{towerthm} for $P$ equal to the perfect sequences of lengths
$1,2,4,8,16$ and $6$ and $a\le7$, and Corollary~\ref{shapecor} in three dimensions for the shapes
$2^a\times4\times2$, $a\le4$, and $2^a\times4\times6$, $a\le3$. In all of these three-dimensional
arrays the AOP for the divisor $2$ held in the earlier-lifted direction as well as in the last; a
later lift multiplies rows of the base by units on both sides, which the AOP of the base does not
obviously survive, and we have no proof that it always does.

\section{Comparison with binary arrays}\label{binsec}

A perfect binary array of shape $M_1\times\cdots\times M_d$ with $M_1\cdots M_d>1$ is equivalent to
a Menon difference set in $\Z_{M_1}\times\cdots\times\Z_{M_d}$, and its size is $4u^2$ for an
integer $u$ \cite[Theorem~3.1]{Jedwab92}. For $2$-power shapes the existence question is settled.

\begin{proposition}[{\cite[Theorem~9.2(i)]{Jedwab92}}]\label{binprop}
Let $a_k\ge1$ for $1\le k\le d$. A perfect binary array of shape $2^{a_1}\times\cdots\times2^{a_d}$
exists if and only if $\sum_ka_k=2y$ is even and $a_k\le y+1$ for all $k$.
\end{proposition}

The sufficiency is Jedwab's recursive construction; the necessity is Kraemer's theorem
\cite{Kraemer93} that an abelian group of order $2^{2y}$ containing a Menon difference set has
exponent at most $2^{y+1}$, which Jedwab states as his Theorem~4.1 and which completed earlier
work of Turyn, Dillon and Davis.

So the binary alphabet admits the shapes $2^n\times2^n$, $2^{n+1}\times2^{n-1}$ and
$2^{n+2}\times2^{n-2}$ in two dimensions and nothing more lopsided; $128\times2$, $64\times4$,
$32\times8\times2$ and every shape of odd total exponent are excluded, as is every array with a
side of odd length other than $1$. Over $\Qe$ Corollary~\ref{shapecor} gives all of these: a
perfect $2^a\times2$ array for every $a$ (the example $p^{(5)}$ above has shape $32\times2$, and
$p^{(7)}$ has shape $128\times2$), a perfect $2^a\times6$ array for every $a$, and so on. The
exponent bound, which over $\{\pm1\}$ ties the longest side of a perfect array to its total size, has
no analogue over the quaternions; and this is the two-dimensional form of the fact, from
\cite{Q8tower}, that Mow's bound on the length of a perfect sequence has no analogue over $\Qe$
either.

We note for comparison that the quaternary alphabet $\{\pm1,\pm i\}$ sits between the two.
Arasu and de Launey \cite{ArasudeLauney01} showed that a perfect quaternary $s\times t$ array is
equivalent to a $(2st,2,2st,st)$-relative difference set in $\Z_s\times\Z_t\times\Z_4$ relative to
the subgroup of order $2$ (their Theorem~13), and constructed perfect quaternary arrays of shape
$2^n3^l\times2^m3^l$ for all $n,m$ with $|n-m|\le4$ (their Theorem~6), so that the aspect ratio
$2^4$ is attained, against $2^2$ for binary arrays; in particular the perfect quaternary sequence
of length $16$ is the case $m=0$, and they conjecture that no perfect quaternary sequence of
greater length exists. Whether $|n-m|\le4$ is also necessary is not settled there, but over
$\{\pm1,\pm i\}$ no perfect array of shape $2^n\times2^m$ with $|n-m|>4$ is known. What removes
every such bound over $\Qe$ is not the size of the alphabet but the doubling of
Theorem~\ref{doublethm}, which has no commutative counterpart.

\section{Single types in higher dimensions}\label{notypesec}

Proposition~5.1 of \cite{Q8tower} states that no sequence of length $M\not\equiv0\pmod4$ over
$\Qe$ has a non-central type, and that after a right unit any sequence of non-central type is
quaternary. The first half of this does not survive in higher dimensions, and it is worth seeing
exactly why.

\begin{proposition}\label{ndnotype}
Let $A$ have shape $M_1\times M'$ over $\Qe$ and type $u\in\{\pm i,\pm j,\pm k\}$ in direction
$1$. Then, after right multiplication by a unit, every entry of $A$ lies in $\{\pm1,\pm i\}$. If
moreover $|M'|=M_2\cdots M_d$ is odd, then $M_1\equiv0\pmod4$.
\end{proposition}
\begin{proof}
Take $u=i$ and write $A=\alpha+\beta j$ with $\alpha,\beta$ complex arrays of disjoint support, and
$E=\E_i[A]=\tilde\alpha+\tilde\beta j$ on $\Z_{4M_1}\times M'$. The relations
$\tilde\alpha_{\mathbf n+(M_1,\mathbf0)}=\tilde\alpha_{\mathbf n}i$ and
$\tilde\beta_{\mathbf n+(M_1,\mathbf0)}=-\tilde\beta_{\mathbf n}i$ put the Fourier transforms
$\hat\alpha,\hat\beta$ on the cosets $f_1\equiv1$ and $f_1\equiv3\pmod4$ of the first frequency
coordinate. The $j$-component of $\theta_E$ vanishes off $\{(qM_1,\mathbf0)\}$ because $A$ has type
$i$, and on that set because $\theta_E(qM_1,\mathbf0)=\sum E_{\mathbf n}(i^*)^qE^*_{\mathbf n}$ is a
sum of group conjugates of $\pm1,\pm i$; so $\hat\alpha(\mathbf f)\hat\beta(-\mathbf f)=0$ for all
$\mathbf f$, the two terms of its transform being supported on different cosets. The complex
component of $\theta_E$ is supported on $\{(qM_1,\mathbf0)\}$, so its transform
$|\hat\alpha(-\mathbf f)|^2+|\hat\beta(-\mathbf f)|^2$ depends only on $f_1\bmod4$, and equals
$16|M'|N_J(A)$ on $f_1\equiv1$ and $16|M'|N_C(A)$ on $f_1\equiv3$. Exactly as in \cite{Q8tower}
this forces one of $\alpha,\beta$ to vanish, and a right unit and a reversal in direction $1$
make $A$ quaternary. This part uses nothing about $M'$.

For quaternary $A$ let $C(\tau_1,\boldsymbol\tau')=\sum_{m<M_1-\tau_1}\sum_{\mathbf n'}
A_{(m,\mathbf n')}A^*_{(m+\tau_1,\mathbf n'+\boldsymbol\tau')}\in\Z[i]$, aperiodic in direction $1$
and periodic in the others. Type $i$ reads $C(\tau_1,\boldsymbol\tau')=i\,C(M_1-\tau_1,
-\boldsymbol\tau')^*$ for $0<\tau_1<M_1$. Each term is a unit of $\Z[i]$, so modulo $\pi=1+i$ the
two sides are $(M_1-\tau_1)|M'|$ and $\tau_1|M'|$, and $(M_1-2\tau_1)|M'|\equiv0\pmod\pi$. If
$|M'|$ is odd this gives $M_1$ even, and then $\tau_1=M_1/2$ with $C=a+bi$ gives $a=b$ and
$a+b\equiv(M_1/2)|M'|\pmod2$, so $4\mid M_1$. If $|M'|$ is even both congruences are empty.
\end{proof}

The second half is sharp. Over $\Qe$ there are no perfect arrays with $|M'|$ odd and $|M'|>1$ in
the ranges we have searched, so the case $|M'|$ odd is in effect the one-dimensional case; and for
$|M'|$ even single non-central types exist in small shapes:
\[
\begin{bmatrix}1&1\\1&-1\end{bmatrix}\ \text{has type }(i,1),
\]
and exhaustive search with $A_{(0,0)}=1$ finds $8$ arrays of shape $2\times2$ of type $i$ in
direction $1$ (and $8$ of type $i$ in direction $1$ with the nega-periodic extension in direction
$2$), $32$ of shape $2\times4$, and none of shape $4\times2$. For shape $1\times M'$ the type
condition says that the sequence $x=A_{(0,\cdot)}$ satisfies $\sum_mx_m\,u\,x^*_{m+\tau}=0$ for
all $\tau\ne0$ and all $u\in\{1,i,-1,-i\}$; there are $2$ such sequences of length $2$ (one is
$[1,i]$), $8$ of length $4$ and none of length $6$. Single types are therefore available in two dimensions, and $(A,Aj)$ is then an $i$-pair in
direction $1$ whenever $N_C(A)=N_J(A)$, but the seed $(P,P)$ of Lemma~\ref{seedlem} is simpler
and always available, and we have not used them.

\section{Relation to the literature}\label{litsec}

\paragraph{Products.} That the Kronecker product of perfect arrays is perfect is classical over
commutative alphabets; \cite[Theorem~1.1.1]{Blake2017} is the product of perfect sequences of
coprime lengths, which folds a two-dimensional product back into a sequence. Over $\Qe$ the
non-commutativity is harmless because the inner correlation sum is a scalar
(Proposition~\ref{kronprop}), and the only care needed is that a left product and a right product
are different arrays. The content of Theorem~\ref{towerthm} relative to products is the family of
perfect arrays $p^{(a)}$, $a\ge5$, which are products in neither order, and the AOP that every
array in the tower carries in the lifted direction.

\paragraph{Arrays over $\Qe$ of unbounded size.} Barrera Acevedo and Jolly \cite{BAJolly14}
obtained perfect arrays over $\Qe$ of unbounded size from Lee sequences, with shapes tied to the
prime from which the Lee sequence is built; the arrays here have shapes $2^{a_1}\times\cdots\times
2^{a_d}$ and $2^{a_1}\times\cdots\times N$, and are given by index functions. Construction~XXIV of
\cite[\S8.4]{Blake2017} gives arrays of shape $2^n\times2^n$ over $\Qe$ by an index function and
was verified for $n<7$; Corollary~\ref{shapecor} proves the existence of these shapes for all $n$,
though by a different construction, and Theorem~\ref{towerthm} with $P$ a perfect sequence of
length $2^n$ gives $2^n\times2^n$ arrays with the AOP for the divisor $2$ in direction $1$.

\paragraph{The recursive array construction over roots of unity.} In \cite{BlakeTower} the
recursion runs in both directions at once, with types in $\Z_K^2$ mixed by the action of
$SL_2(\Z_K)$ through a shear with carry. Here the recursion runs in one direction at a time with
the other directions merely periodic, and the two directions are coupled only through the base
array. Whether there is a genuinely two-dimensional recursion over $\Qe$, in which both directions
carry a non-central twist and are mixed by an operation analogous to the shear, is the question
raised in the concluding section. The structural constraint is that the twists in the several
directions must commute, so a type vector lies in $\{\pm1,\pm u\}^d$ for one non-central $u$.

\paragraph{Williamson matrices and relative difference sets.} By \cite{BADietrich18} a symmetric
perfect sequence over $\Qe$ of length $n$ is equivalent to a $(4n,2,4n,2n)$-relative difference set
in $C_n\times\Qe$, and \cite{BBCN26} enumerate perfect sequences over $\Qe\cup q\Qe$ to length
$21$. The group-theoretic reading of a perfect $M_1\times\cdots\times M_d$ array over $\Qe$ would be
a relative difference set in $C_{M_1}\times\cdots\times C_{M_d}\times\Qe$, and
Corollary~\ref{shapecor} should give such sets for all $2$-power shapes; we have not pursued this.

\section{Conclusions}

The three operations that build perfect sequences over $\Qe$ in \cite{Q8tower}, orthogonality of
commuting twists, the two interleaving lifts and the doubling of $i$-pairs, act on one direction of
an array and leave the others periodic, and the seed $(P,P)$ turns any perfect array $P$ into the
base of a tower in a new direction. The result is perfect arrays over $\Qe$ in every shape
$2^{a_1}\times\cdots\times2^{a_d}$ and $2^{a_1}\times\cdots\times2^{a_{d-1}}\times N$, with index
functions, with the AOP in the last direction lifted, and, from $2^5$ on in that direction, not of
product form. Against this the binary alphabet admits only the shapes allowed by
the exponent bound of Jedwab and Kraemer. The quaternions remove the bound on the aspect ratio of a perfect array
just as, in \cite{Q8tower}, they remove the bound on the length of a perfect sequence, and for the
same reason: the doubling $(x_0,x_1)\mapsto(x_0\il x_1j,\,x_0j\il x_1)$, which re-enters the chain
of twists at $i$ in every length and has no commutative counterpart.

Two questions are left open. The first is whether single arrays of non-central type, which
Proposition~\ref{ndnotype} and the searches of Section~\ref{notypesec} show to exist in two
dimensions, support a construction of their own, and in which shapes they exist. The second is
whether the two directions of an array can be made to interact, as they do over roots of unity in
\cite{BlakeTower}: an $i$-pair in direction $1$ whose members also carry the twists $i$ and $-i$ in
direction $2$ is preserved by the doubling in either direction, since $j$ conjugates $i$ to $-i$ in
both at once, but its lift in direction $1$ has components whose direction-$2$ twists differ, so
the interleaved array is itself an interleaved pair in direction $2$ rather than a typed array. A
calculus of such interleaved pairs would be the quaternion form of the $SL_2$ action of
\cite{BlakeTower}, and might reach shapes that are not products of the one-directional towers.

\section*{Acknowledgements}

I thank Andrew Tirkel for introducing me to quaternionic sequences and arrays almost twenty years
ago.

\bibliographystyle{abbrv}

\end{document}